\documentstyle[12pt,a4wide]{article}

\begin{document}
\begin{flushright}
C-98-60
\end{flushright}
\begin{center}
{\Large\bf The halting problem for universal quantum computers}\\

\vspace {0.5in}
{\bf Tien D. Kieu}\\
CSIRO MST, Private Bag 33, South Clayton 3169, Australia\\
School of Physics, University of Melbourne, Parkville 3052, Australia\\
{\bf Michael Danos}\\
Enrico Fermi Institute, University of Chicago, Chicago, Illinois 60637, USA\\

\vspace{0.5in}

{\bf Abstract}
\end{center}
\begin{quotation}
The halting of universal quantum computers is shown to be incompatible 
with the constraint of unitarity of the dynamics.
\end{quotation}

\vspace{0.5in}
\noindent
The prospects of using quantum dynamics to process information have opened
up new research horizon.  While most of the activities is in the theoretical study and 
experimental realisation of quantum networks~\cite{deutsch1}, the principles --and the 
halting problem, in particular-- of a universal quantum computer~\cite{deutsch} 
(qcomputer for short) are 
still unsettled and have generated many debates recently~\cite{myers, yushi, ozawa, linden}.  
Below we will discuss the general properties and the problem of halting for 
qcomputers.  We then derive
the inconsistencies between the stringent constraint of unitarity of the dynamics and
other desirable properties of the general halting mechanism. 

Let us define a qcomputer and a general mechanism for halting sufficiently precisely for our
discussion purposes.
\begin{itemize}
\item A qcomputer is a closed quantum system governed by a time-independent, unitary 
dynamics $\cal U$ for each time step between the admission of the initial input, which is some vector in the
Hilbert space of $\cal U$ to represent both programs and data, and the final reading-out 
measurements.
\item For halting, it is desirable of the dynamics to be able to
store the output, which is finite in terms of qubit resources, invariantly (that is, 
unchanged under the unitary evolution) after some finite time when the desirable output
has been computed.  This reserved space, from which the output can be read out anytime
afterwards, is mathematically an invariant
subspace $V$ of $\cal U$ which could be defined by the input state itself --so that different input
problems can have different outputs. This is a generalisation of the proposal of qcomputer halting with 
an augmented halt qubit~\cite{deutsch} where the halt qubit is always set at $|0_{\rm h}\rangle$ initially,
and where the final halt state $|1_{\rm h}\rangle$ establishes an invariant subspace --which 
in this case can only signal whether the computation has finished or not.

A qcomputer can hardly be useful without such an invariant subspace.  Unless it is known in advance
--which is not always possible-- when to stop the qcomputer just in time,  
the hard-earned results of computation will be lost in the ever-changing states under the unitary dynamics.

More explicitly,  
\begin{itemize}
\item The initial input state can be as in the special case of the halt qubit or, 
in general,  can be an entanglement of states 
in the invariant subspace and those in its complement,
\begin{eqnarray}
|\psi_0\rangle &=& |0_{\rm h}\rangle\otimes|x_0\rangle + |1_{\rm h}\rangle\otimes|y_0\rangle,
\end{eqnarray}
where $|1_{\rm h}\rangle$ is not restricted to be a state of the halt qubit but can be general and
as large as necessary, but still finite, to store the output.  We employ the block vector and matrix 
notations as follows
\begin{eqnarray}
|0_{\rm h}\rangle\otimes|x_0\rangle &=& \left(\begin{array}{c} 0\\ |x_0\rangle\end{array}\right),\\
|1_{\rm h}\rangle\otimes|y_0\rangle&=&\left(\begin{array}{c} |y_0\rangle\\0 \end{array}\right).
\label{projection}
\end{eqnarray}
In this representation,
\begin{eqnarray}
{\cal U} &=& \left(\begin{array}{cc} A&\alpha\\ 0&B\end{array}\right).
\label{form}
\end{eqnarray}
The null matrix at the lower left corner is a consequence of the desirable invariance of the 
corresponding
subspace.  Note that the vectors $|x\rangle$, $|y\rangle$ and the block matrices above are of infinite 
dimensions in general (as with the dimensions of the classical Turing machine tape).

\item After some finite time $t=N_0$, for and depending on the given input state, the state 
lies entirely in the invariant subspace.  Thereafter, the state remains there.
\begin{eqnarray}
{\rm For} \;\;N<N_0: && \langle 0_{\rm h}|{\cal U}^{N} |\psi_0\rangle = B^N|x_0\rangle 
\not= 0,
\label{1}\\
{\rm for} \;\;N\ge N_0: && \left\{\begin{array}{ll}
{\cal U}^{N} |\psi_0\rangle &= |1_{\rm h}\rangle\otimes|y_N\rangle,\\
\langle 0_{\rm h}|{\cal U}^{N} |\psi_0\rangle &= B^{N}|x_0\rangle = 0.
\end{array} \right.
\label{2}
\end{eqnarray}
\end{itemize}
\end{itemize}

We will now show that the above requirements are not compatible with the stringent constraint of unitarity,
\begin{eqnarray}
{\bf 1} &=& {\cal U}{\cal U}^\dagger = {\cal U}^\dagger{\cal U}, \nonumber\\
\left(\begin{array}{cc} 1&0\\ 0&1\end{array} \right) &=& 
\left(\begin{array}{cc} AA^\dagger + \alpha\alpha^\dagger&\alpha B^\dagger\\ B\alpha^\dagger&BB^\dagger
\end{array} \right) = 
\left(\begin{array}{cc} A^\dagger A & A^\dagger\alpha\\ \alpha^\dagger A&B^\dagger B + \alpha^\dagger\alpha
\end{array} \right).
\label{unitarity}
\end{eqnarray}
If the block matrices are of finite dimensions, expressions~(\ref{result1}) and~(\ref{result2}) below can be 
deduced immediately, from which inconsistencies will unavoidably follow.

For the general infinite (block) matrices, when left hand and right hand inverses can be different or separately
non-existent, we have to take few more steps and make use of the following
two lemmas~\cite{matrix} for an infinite matrix $V$:
\begin{itemize}
\item {\em If $V$ has a right hand (r.h.) inverse $V^{-1}$, then the transpose of $V^{-1}$, $(V^{-1})^t$, is 
a r.h. inverse of $V^t$ in a field [ring] in which $(V^{-1})^tV^t(V^t)^{-1}$ is associative.}
\item {\em If $V$ has a r.h. inverse $V^{-1}$, then the complex conjugate of $V^{-1}$, $(V^{-1})^*$, is a r.h.
inverse of $V^*$, the conjugate of $V$.}
\end{itemize}
Since, from~(\ref{unitarity}), $BB^\dagger=1$ and since the requirement that $B^\dagger BB^\dagger$ is 
associative is a reasonable assumption for a universal reversible qcomputer, we can deduce 
from the two lemmas above that
\begin{eqnarray}
B^\dagger B &=& 1;\label{result1}
\end{eqnarray}
upon which, also from~(\ref{unitarity}),
\begin{eqnarray}
\alpha^\dagger\alpha &=& 0.
\end{eqnarray}
It follows from the last expression that for an arbitrary $|x\rangle$ in the appropriate domain:
\begin{eqnarray}
0 &=& \langle x|\alpha^\dagger\alpha|x\rangle,\nonumber\\
&=& \left|\alpha|x\rangle\right|^2.
\end{eqnarray}
Thus
\begin{eqnarray}
\alpha = 0.
\label{result2}
\end{eqnarray}
In a similar fashion, it can also be shown that $AA^\dagger=1$ and $\alpha^\dagger=0$.

The unitarity result~(\ref{result1}) and the expressions~(\ref{1},\ref{2}) of 
the desirable properties
of halting are clearly incompatible.  Thus there cannot be any switching to {\it any} 
invariant subspace after a finite number of time steps for a general initial input:  {\it the 
qcomputer simply cannot halt}.   

The null result of~(\ref{result2}) also strengthens further our conclusion 
as it implies that the complement subspace would also be invariant under the dynamics of $\cal U$.  
And this once again demonstrates the negative result for halting since any initial entanglement between 
the two subspaces cannot be subsequently removed.  (Many contradictions in the halting mechanism can also
be constructed from expression~(\ref{result2}) which says that the desirable invariant subspace is also
invariant with respect to the reversed quantum dynamics of the qcomputer.  For example,
if the qcomputer is run backward at some time after $N_0$, when the state would have been 
already in the invariant subspace, then the subsequent states would have ever remained there, 
contradicting to the fact that they should not lie entirely in this subspace for a forward
time less than $N_0$.)

In particular and as a special case, the above arguments show that a halt qubit, as proposed in~\cite{deutsch}, 
cannot perform its duty because it can never change its state once it is initially set in one of the 
two states of a measurement basis.  This fact can also be proved independently of the general proof
above.  Let $|\psi_N\rangle$ be the state at the time step $N$.  And let $N_0$ be the switching
time when $|\psi_{N_0-1}\rangle$, having the halt qubit in state $|0_{\rm h}\rangle$, is turned into one in the 
invariant subspace $V$ that has the halt qubit in the state $|1_{\rm h}\rangle$.  Since $|\psi_{N_0-1}\rangle$
is orthogonal to $V$, then ${\cal U}|\psi_{N_0-1}\rangle = |\psi_{N_0}\rangle$ is orthogonal to
the unitary transformation ${\cal U}$ of any vector in $V$.  In particular, $|\psi_{N_0}\rangle$ is orthogonal to the 
unitary transformation of every vector of some orthonormal basis of $V$.  Such a situation can 
only happens if $|\psi_{N_0}\rangle=0$ since $|\psi_{N_0}\rangle$ is in the invariant
subspace $V$ by assumption and the unitary transformation of an orthonormal basis also constitutes
another orthonormal basis.  Because $|\psi_{N_0-1}\rangle \not= 0$ for non-trivial computation, 
we thus arrive at a contradiction proving that the halt state cannot be switched.

In summary, for any legitimate input, unitarity of the dynamics dictates that 
no universal quantum computers can halt in the 
general sense that after a finite number of time steps some acquired results can then be stored unchanged for 
retrieval (via measurements) at an arbitrary time afterwards.  The lack of a working mechanism 
for halting also has implications for the investigation of general quantum computable 
functions.  It should be noted that, however, our arguments and conclusions herein do not cover the case of 
{\it probabilistic} halting.

We wish to thank Alan Head, Robert Lee and Richard Josza for discussions.

\end{document}